\documentclass{elsart}

\usepackage{notation}
\usepackage{graphicx}
\graphicspath{{wander/}{fig/}}
\usepackage{subfigure}
\usepackage{ipeinc}

\usepackage{bluered}
\usepackage{thrmlist}

\let\realendpf=\endpf
\def\endpf{\hspace*{\fill}\qed\realendpf}

\newif\ifabstract
\abstractfalse
\newif\iffull
\ifabstract \fullfalse \else \fulltrue \fi

\begin{document}
\runauthor{Aichholzer, Bremner, Demaine, Meijer, Sacrist\'an, and Soss}
\begin{frontmatter}
\title{Long Proteins with Unique Optimal Foldings in the \protect\HP{} 
Model\thanksref{prelim}}
\thanks[prelim]{A preliminary version of this paper appeared at the 17th European Conference on Computational Geometry~\protect\cite{abdmss-lpuof-01}.}

\author[TUG]{Oswin Aichholzer\thanksref{APART}},
\author[UNB]{David Bremner\thanksref{NSERC}},
\author[MIT]{Erik D. Demaine},
\author[Queens]{Henk Meijer},
\author[UPC]{Vera Sacrist\'an\thanksref{DURSI}}, and
\author[CCG]{Michael Soss}
\address[TUG]{Institut f\"ur Grundlagen der Informationsverarbeitung,
         Technische Universit\"at Graz, Inffeldgasse 16b,
         A-8010 Graz, Austria, \texttt{oaich@igi.tu-graz.ac.at}}
\address[UNB]{Faculty of Computer Science, University of New Brunswick,
         P. O. Box 4400, Fredericton, N. B. E3B 5A3, Canada, email:
         \texttt{bremner@unb.ca}}
\address[MIT]{MIT Laboratory for Computer Science,
         200 Technology Square, \\ Cambridge, MA 02139, USA,
         \texttt{edemaine@mit.edu}}
\address[Queens]{Department of Computing and Information Science,
         Queen's University, Kingston, Ontario K7L 3N6, Canada,
         \texttt{henk@cs.queensu.ca}}
\address[UPC]{Departament de Matem\`atica Aplicada II, Universitat
         Polit\`ecnica de Catalunya, Pau Gargallo 5, 08028 Barcelona,
         Spain, \texttt{vera@ma2.upc.es}.}
\address[CCG]{Chemical Computing Group, 1010 Sherbrooke St.\ West, Suite 910,\\
         Montreal, Quebec H3A~2R7, Canada, \texttt{soss@chemcomp.com}.}

\thanks[APART]{Supported by the Austrian Programme for Advanced Research
               and Technology (APART).}
\thanks[NSERC]{Partially supported by NSERC Canada.}
\thanks[DURSI]{Supported by DURSI Gen.\ Cat.\ 1999SGR00356 and
               Proyecto DGES-MEC PB98-0933.}

\begin{abstract}
  It is widely accepted that (1) the natural or folded state of
  proteins is a global energy minimum, and (2) in most cases proteins 
  fold to a unique state determined by their amino acid sequence.  The
  \HP{} (hydrophobic-hydrophilic) model is a simple combinatorial model
  designed to answer qualitative questions about the protein folding
  process.  In this paper we consider a problem suggested by Brian
  Hayes in 1998: what proteins in the two-dimensional \HP{} model have
  \emph{unique} optimal (minimum energy) foldings?  In particular, we
  prove that there are closed chains of monomers (amino acids) with
  this property for all (even) lengths; and that there are open
  monomer chains with this property for all lengths divisible by four.
\end{abstract}

\end{frontmatter}

\section{Introduction}

Protein folding \iffull \cite{Chan-Dill-1993,Hayes-1998,Merz-LeGrand-1994} \fi
is a central problem in molecular and computational biology
with the potential to reveal an
understanding of the function and behavior of proteins, the
building blocks of life.  Such an understanding would greatly influence many
areas in biology and medicine such as drug design.
\iffull In broad terms, the protein-folding problem is to determine
how proteins so consistently fold into a stable state.  The most
ambitious goal is to understand the entire \emph{folding pathway} (see
e.g.~\cite{sa-umpsp-2001}), i.e.\ the complete dynamics and/or chemical
changes involved in going from an unfolded linear state into a compact
folded state.  Although naturally posed as a numerical simulation,
there are several problems of scale, including the small energy
differences between folded and unfolded states, and the extremely
short interval (approximately $10^{-15}$ seconds) for which the dynamics equations
remain valid, compared to the milliseconds to seconds over which the
folding takes place~\cite{cb-cmb-2000}.  The \emph{thermodynamic
  hypothesis}, first developed by Anfinsen~\cite{anfinsen-1972},
proposes that proteins fold to a \emph{minimum energy} state.  This
motivates the attempt to predict protein folding by solving certain
optimization problems.  There are two main difficulties with this
approach: there is as yet no scientific consensus on what the precise
energy function to be minimized might be, and the functions commonly
used lead to extremely difficult optimization
problems~\cite{fp-occmb-2000,neumaier97molecular}.

\fi

One of the most popular models of protein folding is the
hydrophobic-hydrophilic (\HP{})
model~\cite{Chan-Dill-1993,Dill-1990,Hayes-1998}.
In the \HP{} model, proteins are modelled
as a chains whose vertices are marked either $\H$ (hydrophobic) or
$\P$ (hydrophilic); the resulting chain is embedded in some lattice.
$\H$ nodes are considered to attract each other while $\P$ nodes are
neutral.  An \emph{optimal} embedding is one that maximizes the number
of \red-\red\ contacts. This combinatorial model is attractive in its
simplicity, and already seems to capture several essential features of
protein folding such as the tendency for the hydrophobic components to
fold to the center of a globular (compactly folded) protein
\cite{Chan-Dill-1993}.
Unlike more sophisticated models of protein folding, the main goal of
the \HP{} model is to explore broad qualitative questions about
protein folding such as whether the dominant interactions are local or
global with respect to the chain.  For a nice survey of the kinds of
questions asked and conclusions drawn, see~\cite{dby-ppf-95}.

While the \HP{} model is most intuitively defined in 3D to match the
physical world, it is arguably more realistic as a 2D model for
currently computationally feasible sizes.  The basic reason for this
is that the perimeter-to-area ratio of a short 2D chain is a close
approximation to the surface-to-volume ratio of a long 3D
chain~\cite{Chan-Dill-1993,Hayes-1998}.

Much work has been done on the \HP{}
model\iffull~\cite{Agarwala-Batzoglou-Dancik-Decatur-Farach-Hannenhalli-Muthukrishnan-Skiena-1997,Backofen-1998-CP,Backofen-1998-PSB,Backofen-2000-CPM,Backofen-Will-2001-CPM,Berger-Leighton-1998,Bornberg-Bauer-1997,Chan-Dill-1990,Chan-Dill-1991,Crescenzi-Goldman-Papadimitriou-Piccolboni-Yannakakis-1998,Dill-Fiebig-Chan-1993,Hart-Istrail-1995,Lau-Dill-1989,Lau-Dill-1990,Lipman-Wilber-1991,Unger-Moult-1993a,Unger-Moult-1993b,Unger-Moult-1993c}\fi.
Without theoretical guarantees,
there are many heuristic approaches (e.g.,
\cite{Dill-Fiebig-Chan-1993,Bornberg-Bauer-1997})
and exhaustive approaches (e.g., \cite{Backofen-1998-CP,Backofen-1998-PSB}).
In theoretical computer science, Berger and Leighton
\cite{Berger-Leighton-1998} proved NP-completeness of finding the
optimal folding in 3D, and Crescenzi et al.\ 
\cite{Crescenzi-Goldman-Papadimitriou-Piccolboni-Yannakakis-1998}
proved NP-completeness in 2D\@.
Hart and Istrail~\cite{Hart-Istrail-1995} have developed a
$3/8$-approximation in 3D and a $1/4$-approximation in 2D
of the number of \red-\red\ contacts in the \HP\ model.
Newman~\cite{Newman-2002} just developed a $1/3$-approximation in 2D.
Agarwala et al.~\cite{Agarwala-Batzoglou-Dancik-Decatur-Farach-Hannenhalli-Muthukrishnan-Skiena-1997}
have developed constant-factor approximation algorithms for a generalized \HP\
model allowing multiple levels of hydrophobicity in the 2D triangular lattice
and the 3D face-centered cube (FCC) lattice.

In this paper we are concerned with the question of whether or not
\HP{} chains have \emph{unique} optimal embeddings.  This is a natural
interpretation of the thermodynamic hypothesis in this model, and a
natural model of folding stability.  There are other factors to consider,
e.g.\ \v{S}ali et al.~\cite{ssk-hdpf-94,ssk-kpflm-94} consider a
folding stable if there is a large score gap between it and the next
best folding, but uniqueness seems like a good candidate for making
\HP{} strings more ``protein-like'' for the following reasons (among others):
\begin{itemize}
\item Insisting on uniqueness of optimal embeddings defeats the known
  proofs of NP-hardness~\cite{Berger-Leighton-1998,Crescenzi-Goldman-Papadimitriou-Piccolboni-Yannakakis-1998}, 
\item The \HP{} chains that produce protein-like 3D structures have a
  small number of optimal foldings~\cite{dby-ppf-95}, 
\item Algorithmically it is easy to design an \HP{} chain that folds to
  particular shape~\cite{Kleinberg-1999} as \emph{one} of its optimal
  states, and
\item Experiments have shown that synthetically designed polymers tend to have
  many optimal embeddings, and also not fold stably~\cite{dby-ppf-95}.
\end{itemize}

In particular we explore a problem
suggested by Brian Hayes \cite{Hayes-1998} about the existence of
stable protein foldings of all lengths.  We solve this problem in a
positive sense for circular protein strands.  We also nearly solve the
problem for open strands by exhibiting an infinite class of proteins
having unique optimal foldings.
       More precisely, we prove the following main results, in a sense establishing
the existence of stable protein foldings in the \HP{} model:
\begin{enumerate}
\item We exhibit a simple family of closed chains of monomers, one for
  every possible (even) length, and prove that each chain has a unique optimal
  folding according to the \HP{} model.

\item We exhibit a related family of open chains of monomers, one for every
  length divisible by 4, with the same uniquely-foldable property.  Note that a
  result as strong as (1) cannot be obtained for open chains, because there are
  some lengths for which no uniquely foldable open chains exist.
\end{enumerate}
 \noindent
 In addition, we observe a complementary result about the ambiguity of folding:
 \begin{enumerate}
 \item[3.] We exhibit a family of (open or closed) chains of monomers,
   one for every length divisible by 12, and prove that each chain has
   $2^{\Omega(n)}$ different optimal foldings, each with $\Omega(n)$
   contacts.  In biological terminology, these proteins have a highly
   degenerate ground state~\cite{Hayes-1998}.
 \end{enumerate}


\section{\HP{} Model}
\label{\HP{} Model}

In this section we review the \HP{} model and introduce some
terminology common to the rest of the paper.

Proteins are chains of monomers, each monomer one of the 20 naturally
occurring amino acids.  In the \HP{} model, only two types of monomers
are distinguished: \emph{hydrophobic} (\H), which tend to bundle
together to avoid surrounding water, and \emph{polar} or
\emph{hydrophilic} (\P), which are attracted to water and are
frequently found on the surface of a folding \cite{Chan-Dill-1993}.
In our figures we use small gray disks to denote \H{} monomers and black
disks to denote \P{} monomers.  These monomers are strung together in
some combination to form an \emph{\HP{} chain}, either an open chain
(path or arc) or a closed chain (cycle or polygon).

Proteins are folded onto the regular square lattice.  More formally, a
\emph{lattice embedding} of a graph is a placement of vertices on distinct
points of the (regular square) lattice such that each edge of the graph maps to
two adjacent (unit-distance) points on the lattice.
In the \HP{} model,
proteins must fold according to lattice embeddings, so we also
call such embeddings \emph{foldings}.

The quality of a folding in the \HP{} model is simply given by the
number of hydrophobic monomers (light-gray \redd\ nodes) that are not
adjacent in the protein but adjacent in the folding.  More formally,
the \emph{\bond{} graph} of a folding has the same vertex set as the
chain, and there is an edge between every two \redd\ vertices that are
adjacent in the folding onto the lattice, but not adjacent along the
chain.  The edges of the \bond{} graph are called \emph{\bond{}s}; in
our figures, \bond{}s are drawn as light-gray edges.

\begin{figure}[htbp]
  \begin{center}
\subfigure[\mbox{no \bond{}s}]{\includegraphics{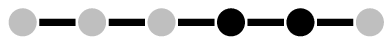}}
\subfiggap{}
\capsubfigure[1 \bond]{\includegraphics{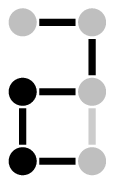}}
\subfiggap{}
\capsubfigure[1 \bond{}]{\includegraphics{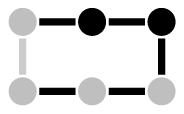}}
\subfiggap{}
\capsubfigure[2 \bond{}s]{\includegraphics{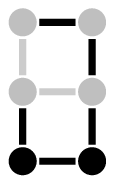}}
\end{center}
  \caption{An optimal folding is one that maximizes the number of \bond{}s.}
  \label{fig:optimal}
\end{figure}
An \emph{optimal} folding maximizes the number of \bond{}s over all
foldings (see Figure~\ref{fig:optimal}).  Intuitively, if a protein is
folded to bring together many hydrophobic monomers (\H{} nodes),
then those monomers are hidden from the surrounding water
as much as possible.


There is a natural bijection between strings in $\{\red, \blue\}^*$ and protein
chains.  We consider the nodes in a chain as labeled by their order in the
string. We sometimes use a limited form of regular expressions to describe
chains where e.g.\ $\red^k$ indicates $k$ \redd\ nodes in sequence.  Similarly,
if we walk along an embedded chain in the order given and read off the
direction of each edge, we can encode foldings as strings in $\{\east, \west,
\north, \south\}^*$.

\begin{figure}[htbp]
  \begin{center}
    \begin{tabular}{@{\extracolsep{1em}}ccc}
  \includegraphics[scale=1.5]{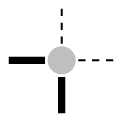} &
  \includegraphics[scale=1.5]{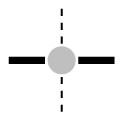}  &
  \includegraphics[scale=1.5]{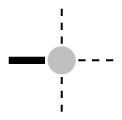}
  \end{tabular}
\end{center}
  \caption{Missing \bond{}s}
  \label{fig:missing}
\end{figure}

For any \sticky{} node $v$ in a lattice embedded chain, consider its
$4$ neighbouring lattice points.  For each of the neighbouring lattice
points that is occupied by neither an adjacent node on the chain, nor by an
\sticky{} node, we call the corresponding lattice edge a \emph{missing
  \bond{}}.  We will also define the number of \bond{}s adjacent to $v$ as
its \emph{\bond{} degree}.  Therefore
 for an endpoint \sticky{} node, a vertex's \bond{}
degree plus its missing \bond{}s totals three;
 for a nonendpoint \sticky{} node, its \bond{} degree plus 
its number of missing \bond{}s is 2.

We will further classify missing \bond{}s into two groups.  Consider
the axis-parallel bounding box of the chain.  If the missing \bond{}
corresponds to an edge outside the bounding box, we refer to it as an
\emph{external missing \bond{}}.  We can further classify an external
missing \bond{} by the one of four walls of the bounding box from
which it emanates.  (At a corner of the bounding box, we consider a
missing \bond{} to emanate from the wall to which the \bond{} is
perpendicular.)  A missing \bond{} which corresponds to an edge inside
the (closed) bounding box we refer to as an \emph{internal missing \bond{}}.

\section{General Observations and Ambiguous Foldings}

In this section we \iffull prove \else describe \fi some basic
structural and combinatorial results about \bond{}s in the \HP{}
model,  in particular establishing that some (nontrivial) chains have
exponentially many optimal foldings.

See also \cite{Backofen-2000-CPM,Backofen-Will-2001-CPM} for upper bounds on
the number of contacts based on the patterns of lattice points occupied by
\redd\ nodes.

\begin{fact}
  A folding of an open chain with $h$ \redd\ nodes has at most
  $h+1$ \bond{}s, and a folding of a closed chain with $h$ \redd\ nodes
  has at most $h$ contacts.
  \label{degree-bound}
\end{fact}
\iffull
\begin{proof}
  The sum of the number of \bond{} and chain edges of any vertex is at most four,
  and every node except possibly the ends has at least two incident chain
  edges.
\end{proof}
\fi

\begin{figure}[htbp]
  \begin{center}
    \includegraphics{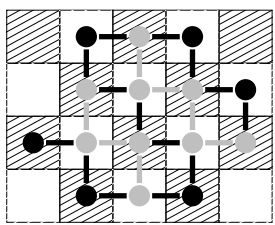}
    \caption{Proving that all lattice-embeddable graphs are bipartite}
    \label{fig:cheq}
  \end{center}
\end{figure}
\begin{fact}
  \label{bipartite}
  Any lattice-embeddable graph is bipartite.
\end{fact}
\iffull
\begin{proof}
  Any subgraph of a bipartite graph is bipartite, and the lattice
  points can be $2$-colored in  checkerboard fashion (see
  Figure~\ref{fig:cheq}). 
\end{proof}
\fi

\begin{cor}
  \label{all-cycles}
  If a folding of a closed chain (or an open chain with \neutral{}
  endpoints) with $h$ \redd\ nodes has $h$ \bond{}s, then its \bond{}
  graph is a union of vertex-disjoint even cycles.
\end{cor}

\iffull
\begin{proof}
   In order to achieve $h$ \bond{}s, every \redd\ vertex must have \bond{}
   degree $2$.
\end{proof}
\fi

\begin{cor}
  \label{red-parity}
  There can be a \bond{} between two \redd\ nodes only if they have opposite
  parity (i.e., there is an even number of nodes between them) in the chain.
\end{cor}
\iffull
\begin{proof}
  The path between two nodes on the chain, along with the \bond{}, form a cycle 
  in the folding.  Thus the result follows from \fref{bipartite}.
\end{proof}
\fi

\iffull
\begin{figure}[htbp]
  \begin{center}
    \includegraphics{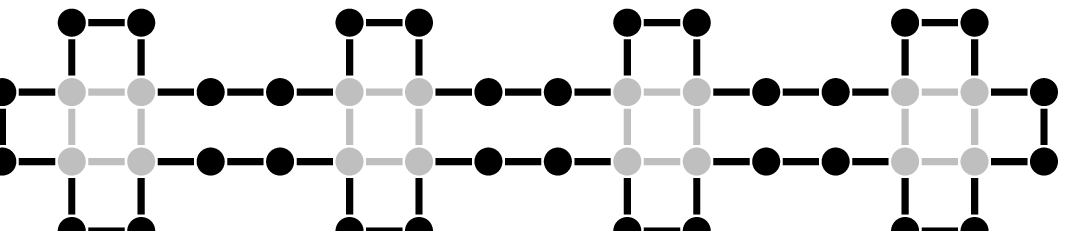}
    \caption{Example of an optimal folding of $(\blue\red\blue)^{4k}$.}
    \label{fig:lineof4cycles}
  \end{center}
\end{figure}
\fi

\iffull
\begin{figure}[htbp]
  \begin{center}
  \begin{tabular}{ccc}
    \includegraphics{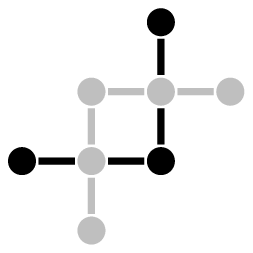} &
\includegraphics{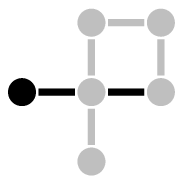} &
\includegraphics{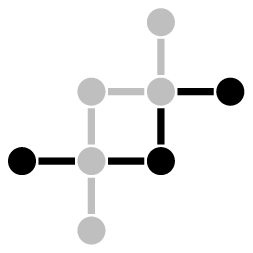} \\
$\H\P\H$ & $\P\H\H$ & $\P\H\P\H$ 
  \end{tabular}
    \caption{Cases for Fact~\protect{\ref{all4cycles}}, with the forbidden 
      subsequences they imply.}
    \label{fig:corners}
  \end{center}
\end{figure}
\fi

\begin{fact}
  \label{all4cycles}
  Any optimal folding of the (open or closed) chain
  $(\blue\red\blue)^{4k}$ has a \bond{} graph consisting of $k$
  $4$-cycles.  \ifabstract (See Figure~\ref{fig:lineof4cycles}.) \fi
\end{fact}
\iffull
\begin{proof}
  $4k$ \bond{}s are achievable e.g.\ by a folding analogous to the one
  shown in Figure~\ref{fig:lineof4cycles}, and no higher number of
  \bond{}s is achievable by \fref{degree-bound}.  By \cref{all-cycles}
  the \bond{} graph is therefore a set of cycles.  Now consider some
  cycle in the \bond{} graph of length greater than 4, and consider
  the leftmost ``$_\ulcorner$'' corner. Up to symmetry, there are
  three cases, illustrated in Figure~\ref{fig:corners}.  In each case
  there is either a singleton \bluee\ or a double \redd\ on the chain.
\end{proof}
\fi


\begin{figure}[htbp]
  \begin{center}
    \begin{tabular}{ccc}
  \raisebox{0.75in}{\includegraphics[scale=1]{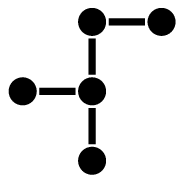}} &  \includegraphics[scale=0.5]{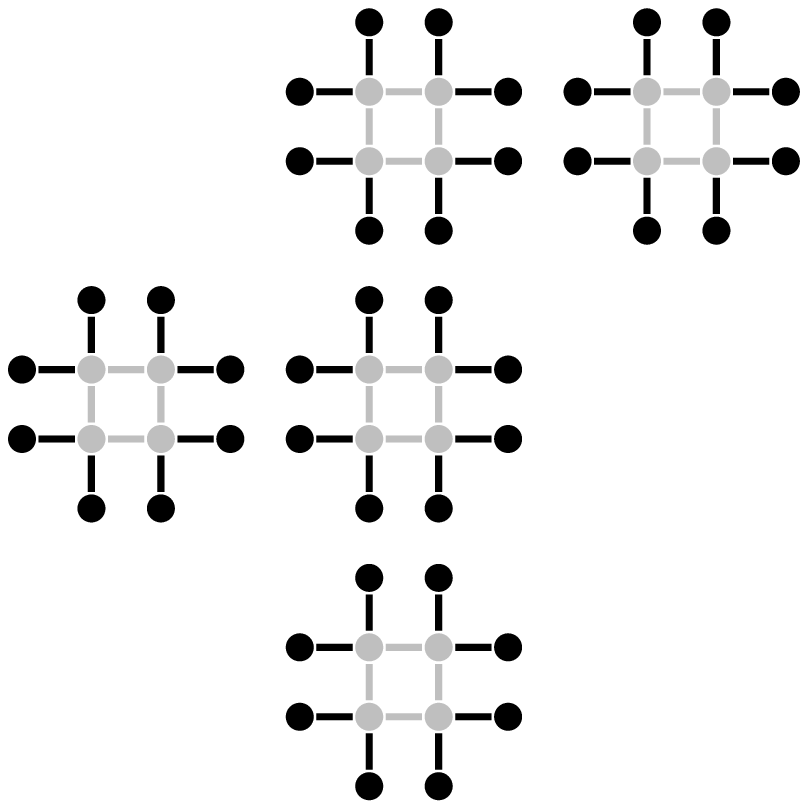} 
  &  \includegraphics[scale=0.5]{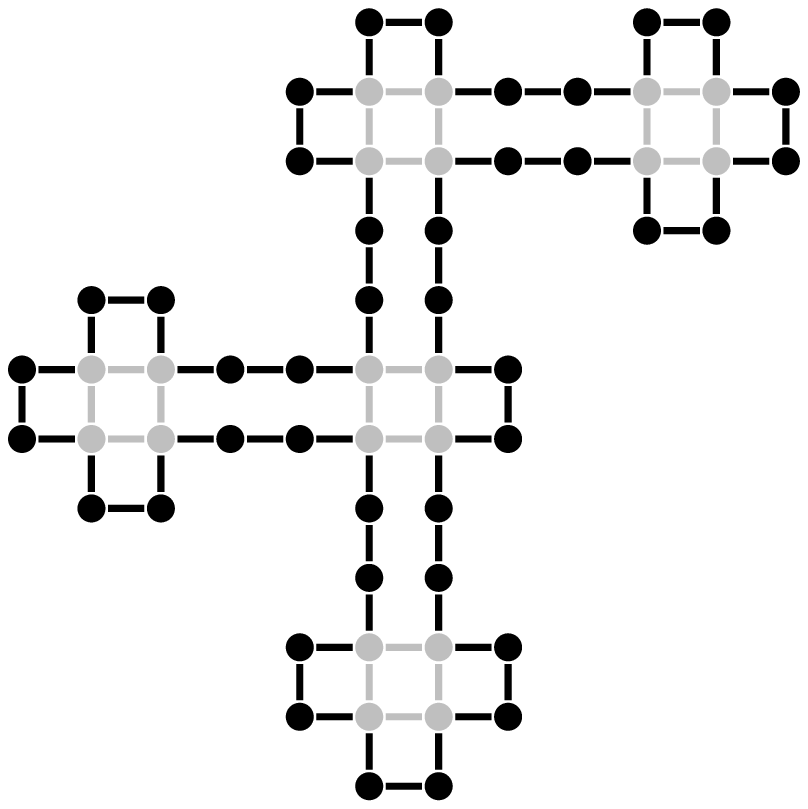}
\end{tabular}
    \caption{Converting a lattice tree into an optimal embedding of $(\blue \red \blue )^{4k}$}
    \label{fig:treeconvert}
  \end{center}
\end{figure}
\begin{fact}
  For any $n=12k$, there exists an $n$-node (open or closed) chain with 
  at least $2^{\Omega(n)}$ optimal foldings, all with
  isomorphic \bond{} graphs of size $\Omega(n)$.
\end{fact}

\iffull
\begin{proof}
  We argue that any lattice-embeddable tree on $k$ nodes corresponds
  to an optimal folding of $(\blue\red\blue)^{4k}$.  To see this
  correspondence, take the embedded tree and scale by 4. Replace each
  node in the tree with a ``gadget'' consisting of a $4$-cycle from
  the \bond{} graph, and the associated forced chain edges (see
  Figure~\ref{fig:treeconvert}).  Finally replace the edges of the
  tree with pairs of edges between adjacent ``gadgets'', and close off
  any remaining $\blue\blue$ pairs with chain edges (in the open-chain
  case, all but one pair is closed off).
  
  Next observe that there are many lattice-embeddable trees on $k$
  nodes.  
 A simple exponential lower bound can be obtained by considering 
 the north/east staircase paths; because there are $2$ choices 
 at each step, this gives a lower bound of $2^k$.
 Each tree (folding) is counted at most a constant number times.
\end{proof}

The preceeding bound on the number of lattice trees can almost
certainly be improved. The number of lattice trees has been studied by
the Statistical Physics community, primarily from the point of view of
deviation from exponential
growth~\cite{ms-saw-96,jvr-ontzd-92,j-elat-2001}.  Based on a
combination of theoretical and experimental results it is
believed~\cite{j-elat-2001} that the number of lattice trees $t_n$
with $n$ nodes obeys the following bound
  \begin{equation*}
    t_n \in \Omega\left ( \frac{3.79^n}{n}  \right ).
  \end{equation*}

\fi

\section{Uniquely Foldable Closed Chains}

In this section we are concerned with closed \HP{} chains whose optimal
foldings are unique (modulo isometries).  For each $k \geq 1$, we
define a closed chain $S_k$ as follows.  Let $A_m$ denote the sequence
$(\rb)^m$.  Define $u = \lceil k/2 \rceil$ and $d = \lfloor k/2
\rfloor$.  Then define $S_k$ as $\blue \, A_u \, \blue \, A_d$.  Note
that $S_k$ has exactly two \emph{\polaredge} edges, i.e.\ edges between two
\neutral{} nodes.
We also define a folding $F_k$ of $S_k$ as follows (see Figure \ref{fig:Sk}).
Let $D_m$ (a ``down staircase'') denote the alternating path $(\east \south)^m$.
Let $U_m$ (an ``up staircase'') denote the alternating path $(\west \north)^m$.
If $k$ is even, define $F_k$ as $\east \, D_d \, \west \, U_u$.
If $k$ is odd, define $F_k$ as $\east \, D_d \, \south \, U_u$.

\iffull
\begin{figure}[htbp]
  \begin{center}
    \begin{tabular}{ccc}
      {\includegraphics{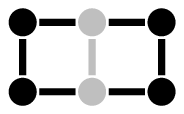}} & {\includegraphics{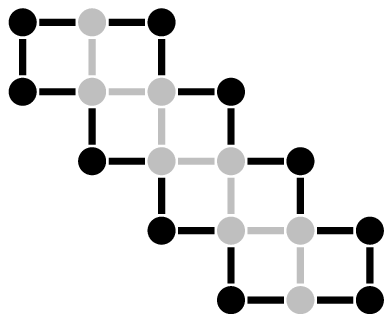}} & {\includegraphics{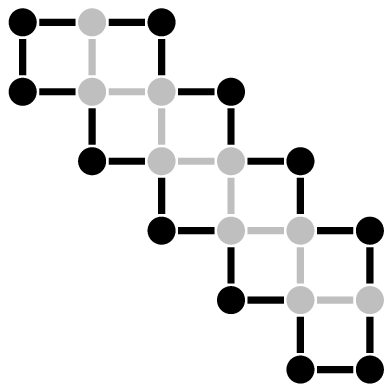}}
    \end{tabular}
    \caption{Examples of $S_k$ folded according to $F_k$ for $k\in \{2, 8, 9\}$.}
    \label{fig:Sk}
  \end{center}
\end{figure}
\fi

\iffull
The main result of this section is the following theorem.
\fi
\begin{theorem}
  \label{sk-unique}
  For each $k\geq 1$, $F_k$ is the unique optimal folding of $S_k$.
\end{theorem}
As well as providing evidence that the \HP{} model captures some
approximation of the mechanism of protein stability for closed chains,
Theorem~\ref{sk-unique} has several less direct consequences.
In the next section we will use this theorem to prove a similar result
about open chains. Furthermore, Theorem~\ref{sk-unique} tells us
something nonobvious about the shape of optimal foldings in the \HP{}
model, namely that there exist (closed) proteins all of whose optimal
foldings are extremely ``nonglobular'' (noncompact).  Along similar lines,
in a preliminary version of this paper~\cite{abdmss-lpuof-01}, we conjectured
that every closed \HP{} chain had an optimal folding with the minimum possible
area (enclosing no grid points).  This conjecture turns out to be
false~\cite{airplane}.

\iffull
\ifabstract The proof has the following outline. \fi

The \emph{conformation graph} of an embedding consists of the
union of the chain edges and the \bond{} graph.  The idea of the proof
of Theorem~\ref{sk-unique} is to show via parity arguments that
the conformation graph of any optimal embedding of $S_j$ is fixed.
Once this is established, the embedding follows from the special form
of the conformation graph (all but one face is a $4$-cycle).

\iffull
\begin{fact}
  \label{path-exists}
  There exists a folding of $S_k$ with $k-1$ \bond{}s, namely $F_k$.
\end{fact}
\fi

\begin{fact}
  The \redd\ nodes of $S_k$ fall into two parity classes, separated into odd
  and even chains by the two \polaredge{} edges.
\end{fact}

In the case of an embedded closed chain $Q$, we distinguish between
\emph{\chordal{}} \bond{}s, i.e.\ those in the interior of $Q$, and 
\emph{\pocket{}} \bond{}s, i.e.\ those exterior to $Q$.

\begin{lemma}
  \label{no-external}
  There are no \pocket{} \bond{}s in an optimal folding of $S_k$.
\end{lemma}
\iffull
\begin{proof}
  Each \redd\ node on the bounding box causes at least one missing
  \bond{}.  From \fref{path-exists}, we know that an optimal folding
  can therefore have at most two \redd\ nodes on the bounding box.
  Because every edge of the chain but two has at least one \redd\ node,
  and there must be at least one edge on each wall of the bounding
  box, there must be at least two \redd\ nodes on the bounding box,
  and this minimum is achievable only if both \polaredge{} edges are on
  distinct edges of the bounding box.  For each \polaredge{} edge construct a
  slab by extending rays perpendicular to the edge from the endpoints
  (see Figure~\ref{fig:slab}).  In order for a \pocket{} \bond{} to
  form, one of the odd or even chains must touch or cross one of the
  slabs.  But if any vertex lies on a slab, the corresponding \polaredge{} edge 
  cannot be on the bounding box.
\end{proof}
\fi

\begin{figure}[htbp]
  \begin{center}
    \includegraphics[trim=72 0 0 0]{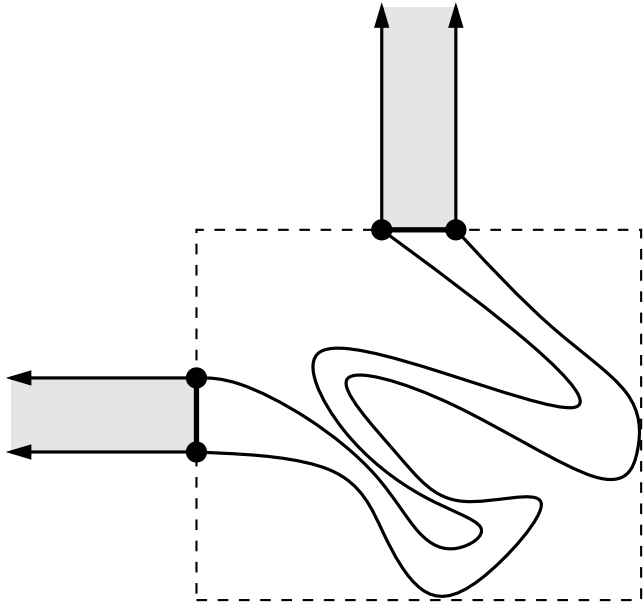}
    \caption{Illustrating the proof of Lemma~\ref{no-external}.}
    \label{fig:slab}
  \end{center}
\end{figure}

\begin{lemma}
  \label{sk-acyclic}
  The \bond{} graph is acyclic in any optimal folding of $S_k$.
\end{lemma}
\iffull
\begin{proof}
  Consider some optimal folding of $S_k$.  By Lemma~\ref{no-external},
  we know that all \bond{}s must be \chordal{}.  Further observe that the
  conformation graph must be a planar graph.  Consider an arbitrary
  planar embedding of the chain $S_k$. Note that each edge of a
  \bond{} cycle must go from the odd chain to the even chain or
  vice-versa.  After two steps along a cycle, there is no way to join
  the first node of the cycle to the (current) last node of the cycle
  without creating a crossing.
\end{proof}
\fi

\begin{cor}
  The \bond{} graph in an optimal folding of $S_k$ is a path with $k$
  nodes and $k-1$ edges.
\end{cor}
\iffull
\begin{proof}
  This follows from~\fref{path-exists} and Lemma~\ref{sk-acyclic}.
\end{proof}
\fi

\begin{figure}[htbp]
  \begin{center}
    \def\IPEfile{pathforce2mono.ipe}\input{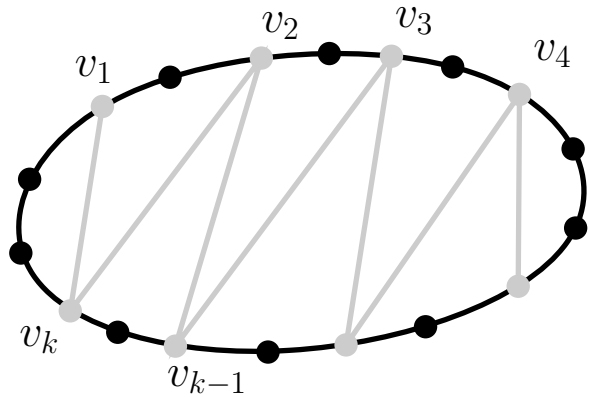}
    \caption{Up to reversal of the labeling, the labeled \bond{} graph of $S_k$ is fixed. }
    \label{fig:pathforce}
  \end{center}
\end{figure}

\begin{lemma}
  \label{fixed-path}
  If $k$ is odd, every optimal folding of $S_k$ has the same labeled
  \bond{} graph.  If $k$ is even, there are two possible labeled
  \bond{} graphs and the mapping from one to the other is given by the
  relabeling $j \mapsto k+1-j$.
\end{lemma}
\iffull
\begin{proof}
  Note that the endpoints of the \bond{} path must be adjacent to one
  of the \polaredge\ edges on the chain because otherwise some \sticky{} node
  would be stranded.  Further note that once the starting point of the
  path is chosen, the rest of the path is determined by an argument
  similar to the proof of Lemma~\ref{sk-acyclic}.  There is no choice
  of starting point for the case of $k$ odd, because both endpoints of
  the \bond{} path must be in the larger parity class of \redd\ nodes.
\end{proof}
\fi

\iffull
We can deduce the following from the proof of Lemma~\ref{fixed-path}.
\fi

\begin{cor}
  In the conformation graph of an optimal folding of $S_k$, 
  \begin{thrmlist}
  \item There are two $4$-cycles of type $\blue\blue\red\red$,
  \item There are $(k-2)$ $4$-cycles of type $\blue\red\red\red$, and
  \item Each \bond{} edge is contained in exactly two $4$-cycles.
  \end{thrmlist}
  \label{four-cycles}
\end{cor}

\iffull
\begin{fact}
  Every $4$-cycle has a unique folding, namely a square.
\end{fact}
\fi

\iffull
\begin{proof}(of Theorem~\ref{sk-unique})
  Consider the folded chain as a polygon, decomposed into
  quadrilaterals by \bond{} edges.  From Corollary~\ref{four-cycles} we
  can see that the dual graph of the decomposition is itself a path.
  We construct the folding by following this dual path.  We start by
  choosing an orientation for one of the $\blue\blue\red\red$
  $4$-cycles and embedding it.  If $k=2$, then we have no choice for the
  final square.  Otherwise, we choose an orientation for the second
  square and embed it.  After the second square, by looking at degrees
  in the \bond{} graph, it follows that we have no choice in embedding
  the next $4$-cycle on the dual path.
  Thus our total choice in embedding was one translation, one rotation, and
  one reflection.
\end{proof}
\fi

\iffull
\begin{cor}
  For every positive even $n$ there is an $n$-node closed \HP{} chain
  with a unique optimal folding.
\end{cor}
\fi

\fi 

\section{Uniquely Foldable Open Chains}

Finally we turn to open \HP{} chains.
Dill et al.~\cite{dby-ppf-95} computed that for chains of length up to
18, about 2\% of chains have unique optimal foldings.  In a similar
vein, Hayes~\cite{Hayes-1998} found that for each $1\leq n \leq 14$
except $3$ and $5$ there is an open chain with a unique optimal
folding.  We have duplicated the experiments of Dill et al.\ and
extended them to chains of length $20$, with (partial) results given
in Table~\ref{tab:percent}.  We have further found experimentally that
there are \HP{} chains with unique optimal foldings for lengths 15 through 25.
Figure~\ref{fig:oswin} illustrates chains with unique optimal foldings for
lengths up to $25$, excluding lengths $2$ (trivial) and $4k$ for $k>3$
(covered by a theorem below).
The bias towards $\H$ nodes is an artifact of our search program, which
enumerates colourings in ``binary-counter'' order with $\H=0$ and
$\P=1$.  It has been reported elsewhere that real proteins have
\sticky{} to \neutral{} ratio of about $2:3$~\cite{Kleinberg-1999}.
We will also see below that in the \HP{} model, unique optimality is
achievable with a ratio very close to $1:1$.

\begin{table}[htbp]
  \begin{center}
    \small
    \begin{tabular}{rrrr}
      $n$ & unique & total & percentage \\\hline
      11 &     65 &     2,048 & 3.174 \\
      12 &     88 &     4,096 & 2.148 \\
      13 &    179 &     8,192 & 2.185 \\
      14 &    387 &    16,384 & 2.362 \\
      15 &    864 &    32,768 & 2.637 \\
      16 &  1,547 &    65,536 & 2.361 \\
      17 &  3,420 &   131,072 & 2.609 \\
      18 &  6,363 &   262,144 & 2.427 \\
      19 & 13,486 &   524,288 & 2.572 \\
      20 & 24,925 & 1,048,576 & 2.377
    \end{tabular}
    \caption{Percentage of \HP{} chains of length $n$ with unique optimal embeddings.}
    \label{tab:percent}
  \end{center}
\end{table}

\iffull
\begin{figure}[htbp]
  \begin{equation*}\arraycolsep=12pt
    \begin{array}{cccc}
  \includegraphics{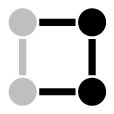} &
  \includegraphics{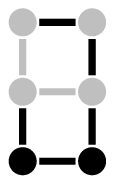} &
  \includegraphics{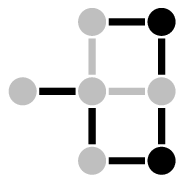} &
  \includegraphics{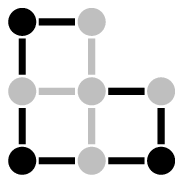}\\
  4 & 6  & 7 & 8\\[3ex]
  \includegraphics{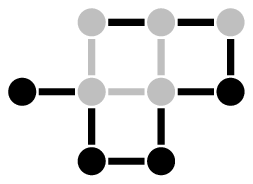}& 
  \includegraphics{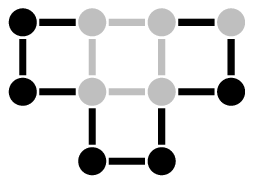} &
  \includegraphics{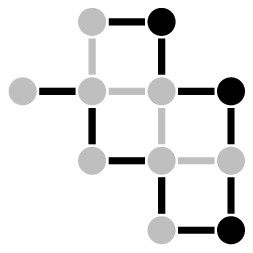} &
  \includegraphics{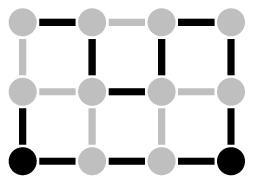}\\
   9  & 10  & 11 & 12\\[3ex]
  \includegraphics{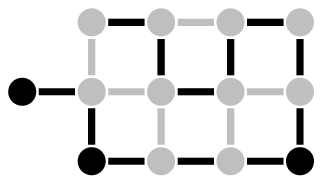} &
  \includegraphics{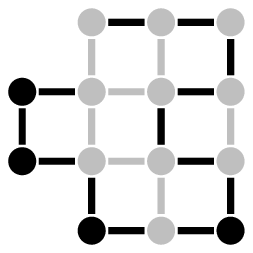} &
\includegraphics{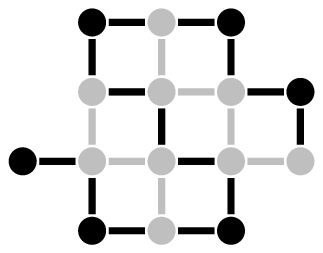} &
\includegraphics{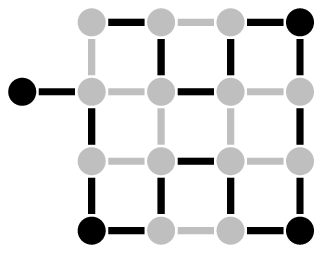} \\
 13 & 14 & 15 & 17\\[3ex]
  \includegraphics[scale=0.8]{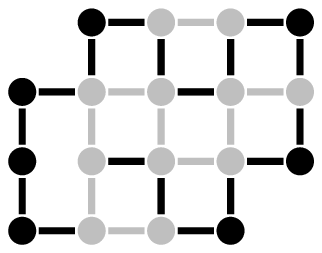} &
  \includegraphics[scale=0.8]{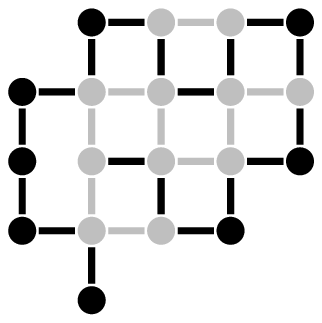} &
  \includegraphics[scale=0.8]{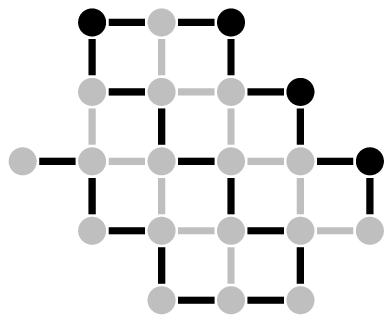} &
  \includegraphics[scale=0.8]{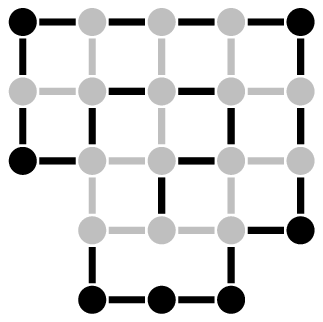}\\
   18 &19 &  21 & 22\\[3ex]
  \includegraphics[scale=0.8]{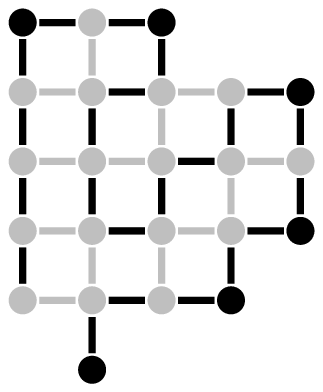} & &
  \includegraphics[scale=0.8]{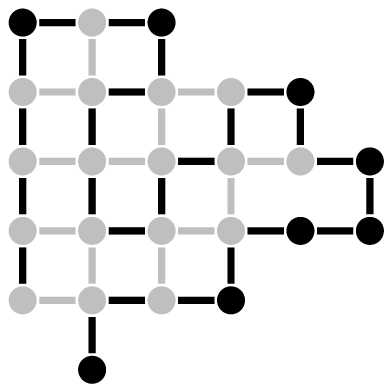} \\
23  & & 25 
  \end{array}
\end{equation*}
\caption{
      Examples of \HP{} chains with unique optimal foldings, shown 
      in their optimal embedding.}
%
      \label{fig:oswin}
\end{figure}
\fi

A natural question is for what values of $n$ there is an $n$-node open chain
with a unique optimal folding.
Based on our results about closed chains,
one approach is to consider the open version of $S_k$ with the first and
last nodes removed.  That is, define $Z_k = (\rb)^u (\br)^d$ where
$u = \lceil k/2 \rceil$ and $d = \lfloor k/2 \rfloor$.
It turns out that this chain has multiple optimal folding for odd $k$,
but only one optimal folding for even $k$ (see Figures~\ref{fig:Z8}
and~\ref{fig:Z9}). 


\iffull
\begin{figure}[htbp]
  \begin{center}
    \includegraphics{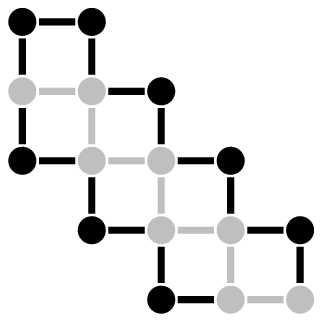}
    \caption{Unique optimal folding of $Z_8$.}
    \label{fig:Z8}
  \end{center}
\end{figure}
\begin{figure}[htbp]
  \begin{center}
    \subfigure{\includegraphics[scale=0.8]{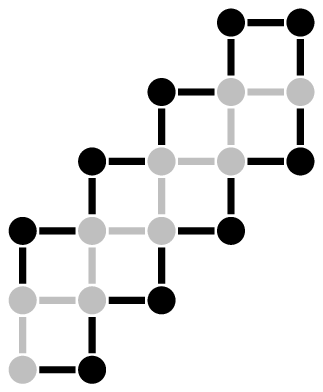}}
    \subfiggap
    \subfigure{\includegraphics[scale=0.8]{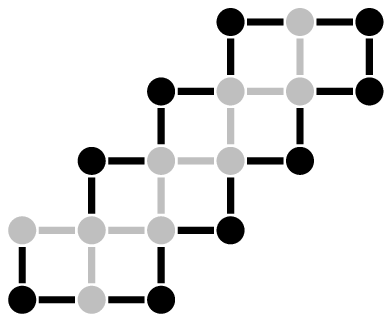}}
    \caption{Two optimal foldings for $Z_9$.}
    \label{fig:Z9}
  \end{center}
\end{figure}
\fi

In what follows, we will establish that up to isometries, the only
optimal embedding of $\ztk$ is what we call the \emph{standard}
embedding, namely the \polaredge{} edge horizontal, the two adjacent edges
down, the remaining edges on the right alternating right and down, and
the remaining edges on the left alternating down and right (the
standard embedding of $Z_8$ is illustrated in \figref{Z8}).  This will establish the following theorem.
\begin{theorem}
  The open chain $Z_{2j} = (\rb)^j (\br)^j$ has a unique optimal embedding for
  each positive $j$.  
\label{theorem:open}
\end{theorem}
Combining this theorem with examples illustrated in
Figure~\ref{fig:oswin}, it turns out that there are open chains with
unique optimal foldings for $n=2$, $n=4$, and $6 \leq n \leq 25$.

Despite the seeming simplicity of the claim, and the similarity to
Theorem~\ref{sk-unique}, the proof of Theorem~\ref{theorem:open}
requires a number of technical lemmas.  We will argue that every
embedding of $\ztk$ has at least four missing external \bond{}s, and
further prove that the standard embedding of $\ztk$ is the only
embedding with exactly four missing \bond{}s.  We accomplish this by
reducing the open-chain case to the closed-chain case discussed in the
previous section.  In particular, we will show that in an optimal
embedding of $\ztk$, the \sticky{} endpoints are on the \bbx{} and
\incontact{}. This will allow us to extend any optimal embedding of
$\ztk$ to an optimal embedding of $S_{2k}$.  The proof can be
summarized as follows:
\begin{enumerate}
\item An optimal embedding has at most $4$ missing \bond{}s (\fref{standard},
  \cref{maxfour}).
\item An optimal embedding has at least $3$ \embs{} and at most one \imb{}
  (\fref{3bb}, Corollaries~\ref{threewalls} and \ref{max1internal}).
\item There are no $\H$ corners (i.e. turns in the chain) on the \bbx{}
  (Lemmas~\ref{nocornercorner} and~\ref{intredcorner}).
\item If an endpoint is on the \bbx{}, either it is \incontact{} with
  the other endpoint, or creates an \imb{} (Lemma~\ref{endpointonbbx}).
\item In an optimal embedding of $Z_{2k}$, there is at most one $\H$
  node on the \bbx{} that is neither a corner, nor an endpoint
  (Facts \ref{solitaryendpoint} and \ref{coupled}, 
   Lemmas \ref{1solitaryendpoint} and \ref{1solitary}).
\end{enumerate}


In the rest of this section, we present the details of the proof of
Theorem~\ref{theorem:open}.  As in the previous section, we start by
observing that any optimal embedding must be at least as good as our
example embedding.
\begin{fact}
  \label{standard}
  The standard embedding of $\ztk$ has only four missing \bond{}s, all external.
\end{fact}

\begin{cor}
  \label{maxfour}
  Any optimal embedding of $\ztk$ has at most four missing \bond{}s.
\end{cor}

  We begin with the following observation.

\begin{fact}
  \label{3bb}
  In any embedding of $Z_{2k}$, either
  \begin{thrmlist}
  \item Three \bbxe{}s contain \sticky{} nodes, and one contains only the 
    \polaredge{} edge of~$\ztk$, or
  \item Four \bbxe{}s contain \sticky{} nodes.
  \end{thrmlist}
\end{fact}
\begin{proof}
  Every \bbxe{} must contain either an edge or an endpoint of $\ztk$.
  Only one edge of $\ztk$ does not contain an \sticky{} node, and each
  endpoint is a \sticky{} node.
\end{proof}

\begin{cor}
  \label{threewalls}
  In an optimal embedding of $\ztk$, there are missing external \bond{}s
  emanating from at least three walls of the bounding box.  The fourth
  wall either contains the \polaredge{} edge, or has a fourth missing external
  \bond{}.
\end{cor}

\begin{cor}
  \label{max1internal}
  Any embedding of $\ztk$ has at least three external missing \bond{}s, and an
  optimal embedding has at most one internal missing \bond{}.
\end{cor}

\begin{fact}
  \label{fact:no-skinny}
  For $k>1$, the \bbx{} of any optimal embedding of $\ztk$ has both
  height and width at least two.
\end{fact}

\begin{proof}
  If either dimension of the \bbx{} is less than 2, then all of the
  \sticky{} nodes are on the \bbx{}.
\end{proof}

\fref{3bb} implies that there is at least one nonendpoint
\sticky{} node $v$ on the bounding box.  We break this down into
two cases.  Two edges adjacent to $v$ could be on the bounding box, in
which case we call $v$ a \emph{straight \sticky{} node}.
Alternatively, only one edge adjacent to $v$ could be on the bounding
box, in which case we call $v$ an \emph{\sticky{} corner}.  
We first argue that in an optimal embedding of $\ztk$ there are no 
\sticky{} corners at the corner of the \bbx{}.

\begin{lemma}
  \label{nocornercorner}
  In an optimal embedding of $\ztk$, there are no \sticky{} corners  at the corner of the \bbx{}.
\end{lemma}
\begin{proof}
  Let $v$ be an \sticky{} corner which is also on the northeast
  corner of the \bbx{}.  We distinguish three cases, as illustrated in
  \figref{redcorner}.  In the first case, $v$ is not adjacent to the
  \polaredge{} edge.  In the second, $v$ is adjacent to the \polaredge{} edge but not on
  the same wall of the \bbx{}.  In the third, $v$ is adjacent to the
  \polaredge{} edge and on the same wall of the \bbx{}.

\begin{figure}[htbp]
  \begin{center}
    \subfigure[]{\label{fig:redcorner1}\includegraphics{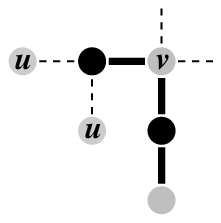}}
    \subfiggap{}
   \subfigure[]{\label{fig:redcorner2}\includegraphics{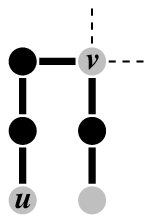}}
\subfiggap{}
   \subfigure[]{\label{fig:redcorner3}\includegraphics{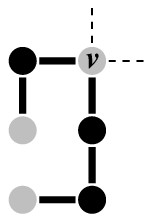}}
   
    \caption{Illustrating the proof of {\protect\lref{nocornercorner}.}}
    \label{fig:redcorner}
  \end{center}
\end{figure}

We first consider the case where $v$ is not adjacent to the \polaredge{}
edge.  Because the two \sticky{} nodes nearest $v$ (two edges away) cannot
both occupy the lattice point southwest from $v$, one of these two
nodes must also be on the \bbx{}.  Suppose there exists an \sticky{}
node on the \bbx{} two edges south from $v$, and consider the position
of the other node $u$ (refer to \figref{redcorner1}).  If $u$ is on
the \bbx{}  then there exists a fourth external missing
\bond{} along only two walls (north and east) of the \bbx{}.  This
implies the presence of a fifth external missing \bond{}.  Thus the
embedding is not optimal by \cref{maxfour}.  The other possibility
is that the path from $v$ to $u$ is west-south, but this creates a
missing internal \bond{} between $u$ and a \neutral{} node.  Because the
presence of a fourth external \bond{} is necessary, this embedding cannot
be optimal.

We next consider the case where $v$ is adjacent to the \polaredge{} edge, but
where the \polaredge{} edge is not on the same \bbx{} wall as $v$ (refer to
\figref{redcorner2}).  This implies that the \polaredge{} edge occupies the
lattice point southwest from $v$, and therefore there exists an
\sticky{} node two edges south from $v$.  Because the east wall has
two external missing \bond{}s and the \polaredge{} edge cannot be on the
south wall, by \cref{threewalls} the \polaredge{} edge must be on the west
wall of the \bbx{} (otherwise, there would be external missing
\bond{}s on each wall, for a total of five, and the embedding would
not be optimal). By Fact~\ref{fact:no-skinny}, this is a
contradiction.


Finally, we consider the case where $v$ is adjacent to the \polaredge{}
edge and on the same wall (see \figref{redcorner3}).  We assume that
the \polaredge{} edge lies directly south of $v$.  Because the north and east
walls have two external missing \bond{}s and the \polaredge{} edge, any
additional external missing \bond{} on these walls would imply
suboptimality by \cref{threewalls}.  Therefore the two \sticky{} nodes
nearest $v$ cannot lie on either of these walls, and must lie as
indicated in the figure.  This causes an internal missing \bond{} between
an \sticky{} node and a \neutral{} node.  Thus there are at least five missing
\bond{}s, and the embedding is not optimal.
\end{proof}

We now expand the restriction on \sticky{} corners to include the
entire \bbxe{}.

\begin{lemma}
  \label{intredcorner}
  There are no \sticky{} corners on the \bbx{} in an optimal embedding
  of~$\ztk$.
\end{lemma}

\begin{proof}
  By Lemma~\ref{nocornercorner}, we need only consider the case
of an \sticky{} corner in the relative interior of a \bbx{} edge. 
Let $v$ be such an \sticky{} corner. We assume that the edges of the
  chain adjacent to $v$ are to the west and the south.  We distinguish
  two cases, as illustrated in \figref{intredcorner}.  In the first
  case, $v$ has no \bond{}s and thus has a missing external \bond{} and a
  missing internal \bond{} (along the wall of the \bbx{}).  In the second
  case, $v$ has an internal \bond{}, which must be with a second
  \sticky{} corner or an endpoint.

\begin{figure}[htbp]
  \begin{center}
    \subfigure[]{{\includegraphics[trim=0 -4 0 0 ]{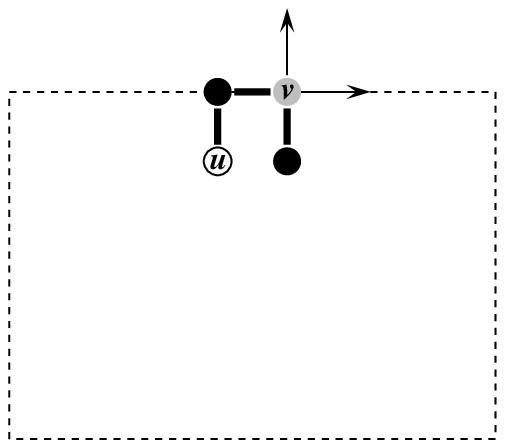}}}
    \subfiggap{}
    \subfigure[]{\includegraphics{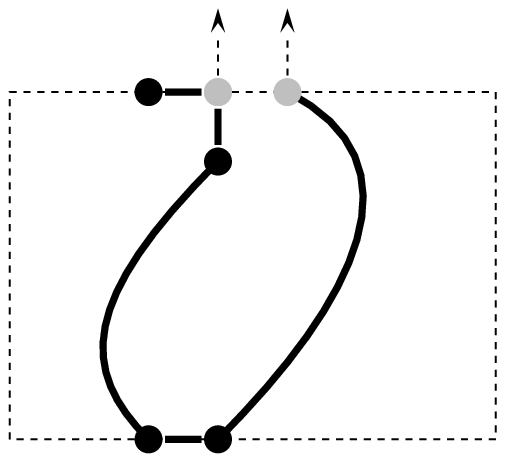}}
    \caption{Illustrating the proof of \protect\lref{intredcorner}.}
    \label{fig:intredcorner}
  \end{center}
\end{figure}

We first consider the case where $v$ has no \bond{}s.  Because there is a
missing internal \bond{}, there can only be three external missing \bond{}s
if the embedding is to be optimal.  Therefore the \polaredge{} edge must be
on the \bbx{}, and cannot be on the north wall.  Furthermore, no other
\sticky{} node can be on the north wall.  
Consider the edge adjacent to $v$ on the north wall, and the node $u$
which follows it.  Because $u$ cannot be on the north wall, its only
possible position is south west of $v$.  This creates a second
internal missing \bond{} between $u$ and a \neutral{} node (if $u$ is an
\sticky{} node) or an extra \emb{} (if $u$ is part of the \polaredge{} edge),
causing suboptimality.

We finally consider the case where $v$ has an internal \bond{}, which must
be with a second \sticky{} corner or endpoint $w$.  Because $w$ has a
second external missing \bond{} on the north wall, the \polaredge{} edge must
lie on the \bbx{} by \cref{threewalls}.  Furthermore, the \polaredge{} edge
must lie on the path from $v$ to $w$ because $v$ and $w$ have
different parity.  Thus the path from $v$ to the \polaredge{} edge creates a
barrier between all points west of this path and all points after the
\polaredge{} edge.  The endpoint of the same parity as $v$ must lie west of
the path, as the chain cannot go outside the \bbx{}.  This endpoint
has three missing \bond{}s, for a total of five for the chain.  Therefore
the embedding is not optimal.
\end{proof}

We have now established that if an \sticky{} node is on the \bbx{}, it
must either be straight or an endpoint.  With respect to endpoints, we
observe the following:
\begin{lemma}
  \label{endpointonbbx}
  In an optimal embedding of $\ztk$, if an endpoint is on the \bbx{},
  then either there is an internal missing \bond{}, or 
  the two endpoints are \incontact{}.
\end{lemma}
\begin{proof}
  An endpoint has three potential \bond{}s; at least one of which lies
  along the wall of the \bbx{}.  Either this is an internal missing
  \bond{}, or the endpoint is adjacent to an \sticky{} node.  Because there
  are no \sticky{} corners in an optimal embedding, this \sticky{}
  node must be the second endpoint.
\end{proof}

\begin{figure}[htbp]
  \begin{center}
    \subfigure[]{\includegraphics{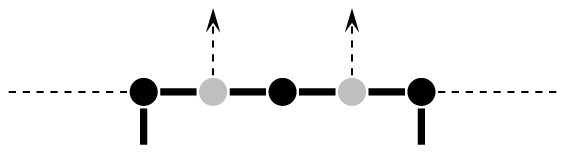}}
    \subfigure[]{\includegraphics{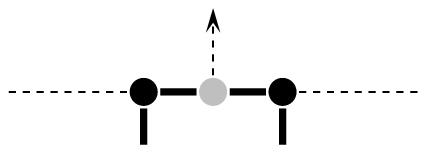}}
    \caption{A pair of coupled straight \sticky{} nodes, and a solitary straight \sticky{} node.}
    \label{fig:coupledsolitary}
  \end{center}
\end{figure}

Because there are only two endpoints and one \polaredge{} edge, we know there
must be at least one straight \sticky{} node on the \bbx{}.  We define
two kinds of straight \sticky{} nodes.  We say a straight \sticky{}
node $v$ is \emph{coupled} if the preceding or following \sticky{}
node is also on the same wall of the \bbx{} as $v$; otherwise it is
\emph{solitary}.  These cases are illustrated in
\figref{coupledsolitary}.

\begin{figure}[htbp]
  \begin{center}
    \includegraphics{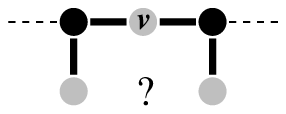}
    \caption{Illustrating the proof of \protect\fref{solitaryendpoint}.}
    \label{fig:solitaryendpoint}
  \end{center}
\end{figure}

\begin{fact}
  \label{solitaryendpoint}
  In an optimal embedding of $\ztk$, a solitary straight \sticky{}
  node must either be adjacent to the \polaredge{} edge or \bond{} with an
  endpoint.
\end{fact}
\begin{proof}
  Let $v$ be a solitary straight \sticky{} node on the north wall of
  the \bbx{} which is not adjacent to the \polaredge{} edge.  Then there must
  be two \sticky{} nodes immediately southwest and southeast of $v$,
  as illustrated in \figref{solitaryendpoint}.  The lattice point
  south of $v$ cannot be the adjacent \neutral{} node of any of the
  \sticky{} nodes, nor can it be vacant, because there would be at least
  two missing internal \bond{}s, violating optimality.  
\end{proof}

\begin{fact}
  In an optimal embedding of $\ztk$, there is at most one pair of
  coupled straight \sticky{} nodes.
  \label{coupled}
\end{fact}
\begin{proof}
  Each $m$-tuple of coupled \sticky{} nodes causes $m$ external
  missing \bond{}s on the same bounding box wall.  There can be at
  most one wall with two external missing \bond{}s, and none with
  three or more.
\end{proof}

\begin{figure}[htbp]
  \begin{center}
    \includegraphics{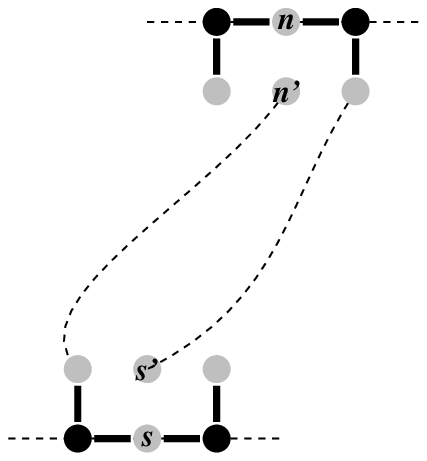}
    \caption{Illustrating the proof of \protect\lref{1solitaryendpoint}.}
    \label{fig:1solitaryendpoint}
  \end{center}
\end{figure}

The following two lemmas establish that there is at most one solitary
straight \sticky{} node.

\begin{lemma}
  \label{1solitaryendpoint}
  In an optimal embedding of $\ztk$, there is at most one solitary
  straight \sticky{} node \incontact{} with an endpoint.
\end{lemma}
\begin{proof}
  Assume there are two such straight \sticky{} nodes, $n$ and $s$,
  which \bond{} with two endpoints, $n'$ and $s'$, respectively.  We
  assume $n$ is on the north wall and $s$ on the south; the argument
  can be slightly modified for any other case.  Because $n'$ and $s'$
  are of different parity, so too must be $n$ and $s$.  Therefore
  there is a path on the chain from $n$ to $s'$ which does not pass
  through $s$, and a path from $s$ to $n'$ which does not pass through
  $n$.  Assume without loss of generality that
  the path from $n$ to $s'$ leaves $n$ to
  the east, as in \figref{1solitaryendpoint}.  Then the path from $s$
  to $n'$ must leave $s$ to the west to avoid intersection.  Because the
  chain is connected, there must be a path from $n$ to $s$.  The only
  possibility left is that the path leaves $n$ to the west, and enters
  $s$ from the east.  However, this requires the path to either leave
  the \bbx, or intersect the rest of the chain, neither of which is
  possible.
\end{proof}

\begin{figure}[htbp]
  \begin{center}
    \subfigure[]{\label{fig:1solitary-a}\includegraphics{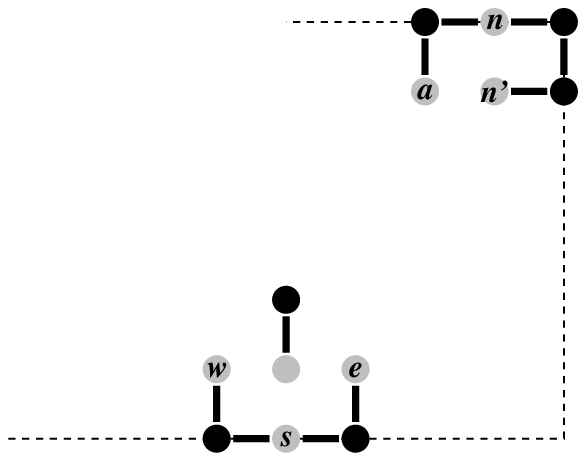}}
    \subfiggap{}
    \subfigure[]{\label{fig:1solitary-bottom}\includegraphics{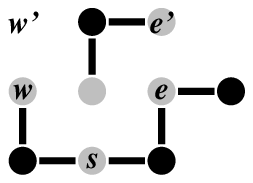}}
    \subfigure[]{\label{fig:1solitaryturn}\includegraphics{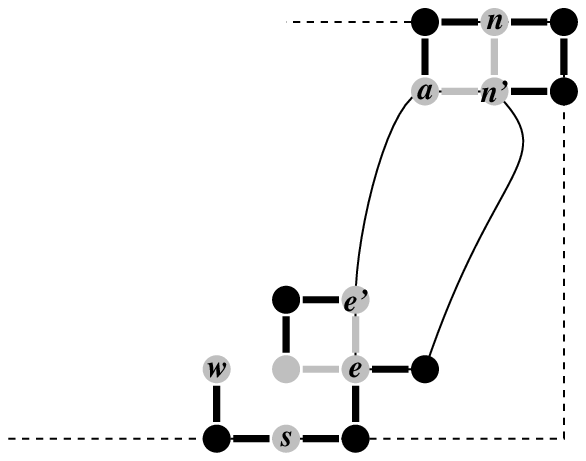}}
    \caption{Illustrating the proof of \protect\lref{1solitary}.}
    \label{fig:1solitary}
  \end{center}
\end{figure}

\begin{lemma}
\label{1solitary}
  In an optimal embedding of $\ztk$, there is at most one solitary straight \sticky{} node on the \bbx{}.
\end{lemma}
\begin{proof}
  Suppose there is more than one straight \sticky{} node on the
  \bbx{}. By Fact~\ref{solitaryendpoint} and Lemma~\ref{1solitaryendpoint}
  there must be exactly two: one must be adjacent to the \polaredge{} edge;
  the other must \bond{} with an endpoint.  We observe that the \polaredge{}
  edge must be on the \bbx{}, because each wall not containing a
  solitary \sticky{} node must otherwise contain either a pair of
  coupled straight \sticky{} nodes or an endpoint; any combination of
  these two possibilities leads to a total of at least $5$ missing
  \bond{}s (apply Lemma~\ref{endpointonbbx} in the case of an endpoint).

  Assume the north, east, and south walls are covered by the two
  solitary straight \sticky{} nodes and the \polaredge{} edge.  We assume the
  configuration in \figref{1solitary-a} without loss of generality
  (the argument here will not depend on whether the two solitary nodes
  are on opposite walls of the \bbx{}).  Call the \sticky{} node on
  the south wall $s$, and the \sticky{} nodes preceding and following
  (northeast and northwest of $s$) $e$ and $w$.  One endpoint of
  the chain is \incontact{} with $e$, $w$, and $s$.  
  
  Consider the possibilities for covering the west wall of the \bbx{}:
  there is either a pair of coupled straight nodes, or an endpoint.  By
  Lemma~\ref{endpointonbbx}, either case results in two missing \bond{}s,
  both either contained in or emanating from the west wall.  It
  follows that we need only find one more missing \bond{} to establish
  suboptimality.

  Consider the \sticky{} node $e$ east of $s$; $e$ cannot be on the
  \bbx{} as this would create an \emb{}.  Placing an \sticky{} node
  east of $e$ just creates another \imb{}, which in turn cannot be
  blocked without creating a missing \bond{} with the \neutral{} node
  between $s$ and $e$ on the chain.  It follows that the chain must
  turn east at $e$; in order to avoid an \imb{}, there must be an $\H$
  node north of $e$ (at $e'$ in Figure~\ref{fig:1solitary-bottom}),
  and a chain edge west from this $\H$ node.  Note that lattice point
  north of $w$ (i.e.~$w'$) cannot contain an \sticky{} node, as this
  would create an \imb{}.
  
  By planarity, the subchains containing $e$ and $e'$ must be
  connected to the \polaredge{} edge as shown in
  Figure~\ref{fig:1solitaryturn}.  Consider the polygon $Q$ formed
  these connecting chains, along with the \bond{}s $ee'$ and $an'$.
  By an argument similar to Lemma~\ref{sk-acyclic} (with the extension
  that no \bond{}s can be formed with $w$), all of the remaining
  \bond{}s for \sticky{} nodes on this polygon must be internal to
  $Q$. Because there is an imbalance in the parity of potential \bond{}s
  for nodes on $Q$, this forces an \imb{}.
%
\end{proof}

We are finally ready to characterize the intersection of 
an optimal embedding of $\ztk$ with its \bbx{}.

\begin{lemma}
  In an optimal embedding of $\ztk$, the two endpoints are on the
  \bbx{} and \incontact{}, and the \polaredge{} edge is adjacent to a
  solitary \sticky{} node, both of which are also on the \bbx{}.
  Furthermore, there are no internal missing \bond{}s.
\end{lemma}

\begin{proof}
  By \lref{1solitary} there is at most one solitary straight \sticky{}
  node on the \bbx{}, and there is of course only one \polaredge{} edge.  On
  the other two walls of the \bbx, there are either two endpoints, or
  one endpoint and one pair of coupled straight \sticky{} nodes. In
  the last case, the endpoint causes one internal missing \bond{} by
  \lref{endpointonbbx}, and the coupled straight \sticky{} nodes cause
  two external missing \bond{}s.  Therefore there are no
  coupled straight \sticky{} nodes.  Furthermore, because both
  endpoints are on the \bbx{}, they must be \incontact{} by
  \lref{endpointonbbx} to avoid having two internal missing \bond{}s.
  Because the two endpoints \bond{}, and are both on the \bbx{}, they
  have two external missing \bond{}s on the same wall.  Therefore there
  are four external missing \bond{}s in an optimal embedding, and thus no
  internal missing \bond{}s.
\end{proof}

\begin{figure}[htbp]
  \begin{center}
    \includegraphics[width=2.5in]{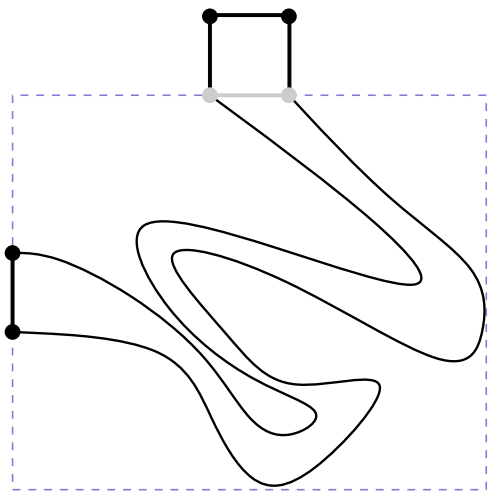}
    \caption{Converting an optimal embedding of $\ztk$ into an optimal embedding into an optimal embedding of $S_{2k}$. }
    \label{fig:convert}
  \end{center}
\end{figure}
The previous lemma claims in essence that the open chain $\ztk$
behaves just like a closed chain in any optimal embedding.  We formalize 
this intuition as follows:
\begin{theorem}
  There are as many optimal embeddings of $S_{2k}$ as there are of 
  $\ztk$.
\end{theorem}
\begin{proof}
  By the preceding lemma, in an optimal embedding the endpoints
  $\ztk$ are \incontact{} and on the \bbx{}; thus we can convert an optimal
  embedding of $\ztk$ into an optimal embedding of $S_{2k}$ by adding 
  a second \polaredge{} edge, outside the \bbx{} (see Figure~\ref{fig:convert}).
\end{proof}
Theorem~\ref{theorem:open} is a straightforward consequence of the
preceding theorem and Theorem~\ref{sk-unique}.

\medskip

We expect that by similar methods we can prove the following:
\begin{conjecture}
  For odd $k \geq 5$, the open chain $Z_k$ has exactly two optimal
  embeddings.
\end{conjecture}
We have computationally verified this conjecture for chains of length up to
$26$, that is, for odd $5 \leq k \leq 13$.
For $k=1$ and $k=3$, $Z_k$ in fact has a unique optimal folding.

\section{Conclusions and Directions for Future Work}

In this paper we considered a natural characterization of stable
protein folding in the 2D \HP{} model, namely uniqueness of optimal
folding.  We established that
\begin{enumerate}
\item There exist closed \HP{} chains with unique optimal folds for
  all (even) lengths, and
\item There exist open \HP{} chains with unique optimal folds for all lengths
  divisible by $4$.
\end{enumerate}
We further observed that
\begin{enumerate}\setcounter{enumi}{2}
\item There exist arbitrarily long \HP{} chains with linear sized
  contact graphs and an exponential number of optimal foldings.
\end{enumerate}

There are several natural directions for future work, involving
more general lattices, asymptotic bounds, and algorithmic questions,
as summarized below.

There is a great deal of natural skepticism about the biological relevance
of results stemming from the 2D square lattice.  While certain
qualitative properties are independant of the lattice
used~\cite{dby-ppf-95}, in the case of the present work it seems clear
that the bipartiteness of the lattice plays an important (and
difficult-to-motivate) role.  It is thus important to consider
nonsquare lattices in 2D, and preferably nonbipartite lattices in 3D.
Examples include the triangular lattice in 2D, and
the face-centered cubic (FCC) lattice suggested by
Neumaier~\cite{neumaier97molecular} as a minimal approximation of
chemical bond distances and angles.
For example,
Agarwala et al.~\cite{Agarwala-Batzoglou-Dancik-Decatur-Farach-Hannenhalli-Muthukrishnan-Skiena-1997}
consider both of these lattices.

The existence of \HP{} chains with unique embeddings in a given
lattice is only a first step to understanding the behaviour of these
models with respect to uniqueness.  Experimental results~\cite{dby-ppf-95} have
shown that about $2\%$ of open chains up to length $18$ have unique optimal
foldings.  It would be very nice to have asymptotic bounds for the
fraction of \HP{} chains with unique optimal foldings. Rather than
considering arbitrary \HP{} chains, it would also be useful to see
what fraction of the proteins in the Protein Data Bank~\cite{PDB} fold
uniquely in the \HP{} model.  It is likely that the best that can be
hoped for is that real proteins have a small number of optimal
foldings.

From a sequence-design point of view, the more interesting question is
not whether there exists an \HP{} sequence with a small number of optimal
foldings, but how to design sequences with this property.  From a
combinatorial point of view, this asks for a characterization of what
sequences have unique (or a small number of) optimal foldings.

From an algorithmic point of view, there are two natural questions.
The first problem is whether there is an efficient algorithm to recognize
sequences with unique optimal foldings.
The second prolem is whether the
problem of finding a minimum energy folding of an \HP{} chain is still
NP-hard when restricted to chains with unique optimal foldings.

Finally, there may be better definitions of folding stability in the
\HP{} model.  We have mentioned the notion of a large gap in the
number of contacts between the optimal folding and the next best
folding~\cite{ssk-hdpf-94,ssk-kpflm-94}.  Given that the \HP{} model
is only approximate, it may also be inappropriate to distinguish
between chains having e.g.~1 and 2 optimal embeddings.


\begin{ack}
The authors would like to thank Godfried Toussaint for pointing us to this
topic and for organizing the workshop at which the research was initiated.
  The
authors would also like to thank Vida Dujmovi\'c, Jeff Erickson, Ferran
Hurtado, and Suneeta Ramaswami for stimulating conversations on this and other
topics, and to thank Tom Shermer and Joe Horton for helpful discussions on 
lattice trees.
\end{ack}

\bibliographystyle{abbrv}
\bibliography{authors,venues,biology,lattice}

\end{document}